\documentclass[times]{article}
\usepackage{isita2006}
\usepackage{epsfig}
\usepackage{mathrsfs}
\usepackage{stmaryrd}
\usepackage{amsmath,amstext,amsfonts,amssymb}
\usepackage{epsfig,float}
\usepackage{graphicx}
\usepackage{cite}
\usepackage{psfrag}
\usepackage{subfigure}
\usepackage{url}
\usepackage{shadow,color,pifont,times,rotate}
\DeclareMathAlphabet{\mathpzc}{OT1}{pzc}{m}{it}

\newcommand{\mb}{\mathbf}

\newcommand{\mc}{\mathcal}

\newtheorem{theorem}{Theorem}[section]
\newtheorem{lemma}[theorem]{Lemma}

\newtheorem{corollary}[theorem]{Corollary}

\title{Outage Behavior of Discrete Memoryless Channels\\ Under Channel Estimation Errors}
\name{Pablo Piantanida$^{\dagger}$, Gerald Matz$^{\ddagger}$, and Pierre Duhamel$^{\dagger}$
\thanks{This work was supported by the European Union IST network of excellence NEWCOM.}}

\address{
\begin{tabular}{cc}
\begin{tabular}{c}
$^{\dagger}$
Laboratoire des Signaux et Syst\`emes\\
CNRS/Sup\'{e}lec\\
F-91192 Gif-sur-Yvette, France\\
E-mail: \{piantanida,pierre.duhamel\}@lss.supelec.fr
\end{tabular}
\begin{tabular}{c}
$^{\ddagger}$
Institute of Communications and Radio-Frequency Engineering\\
Vienna University of Technology\\
A-1040 Wien, Austria\\
E-mail: g.matz@ieee.org
\end{tabular}
\end{tabular}\vspace{-1mm}
}

\begin{document}
\maketitle
\sloppy

\begin{abstract}
Classically, communication systems are designed assuming perfect channel state information at the receiver and/or transmitter. However, in many practical situations, only an estimate of the channel is available that differs from the true channel. We address this channel mismatch scenario by introducing the notion of estimation-induced outage capacity, for which we provide an associated coding theorem and its strong converse, assuming a discrete memoryless channel. The transmitter and receiver strive to construct codes for ensuring reliable communication with a quality of service (QoS), in terms of achieving a target rate with small error probability, no matter which degree of accuracy channel estimation arises during a transmission. We illustrate our ideas via numerical simulations for transmissions over Ricean fading channels using rate-limited feedback channel and maximum likelihood (ML) channel estimation. Our results provide intuitive insights on the impact of the channel estimate and the channel characteristics (SNR, Ricean K-factor, training sequence length, feedback rate, etc.) on the mean outage capacity. 
\end{abstract}

\section{Introduction}

Channel uncertainty, caused e.g.~by time variations/fading, interference, or channel estimation errors, can severely impair the performance of wireless systems. Even if the channel is quasi-static and interference is small, uncertainty induced by imperfect channel state information (CSI) remains. This motivates us to study the design of communication systems which require to ensure information transmission at a target rate satisfying a quality of service (QoS), i.e. reliable communication, no matter which degree of accuracy channel estimation arises during the communication. 

We first review the model for communication under channel uncertainty over a 
% memoryless channel 
discrete memoryless channel (DMC)
with finite input alphabet $\mathscr{X}$ and output alphabet $\mathscr{Y}$ \cite{lapidoth-resume}.  A specific instance of the unknown channel is characterized by a transition probability mass (PM) $W(\cdot|x,\theta)\in\mc{W}_{\Theta}$ with an unknown channel state $\theta\in \Theta\subseteq \mathbb{C}^d$. Here, $\mc{W}_{\Theta}=\big\{W(\cdot|x,\theta)\!: \, x\in \mathscr{X},\,\theta\in \Theta\big\}$ is a family of conditional transition PMs on $\mathscr{Y}$, parameterized by a vector $\theta\in \Theta$. By considering a composite channel model, through the notion of reliable communication based on the average of the error probability over all channel estimation errors. Capacity bounds for additive white Gaussian noise (AWGN) channels with MMSE channel estimation, i.e. imperfect CSI at the receiver (CSIR), and no CSI at the transmitter (CSIT) were derived in \cite{medard-2000}.

Throughout the paper we assume that the channel state, which neither the transmitter nor the receiver knows
exactly, is constant within blocks of duration $T$ symbol periods (coherence time), 
and channel states in different blocks are i.i.d.~$\theta\sim\psi(\theta)$. 
The extension of the DMC $W(\cdot|x,\theta)$ to $n$ channel uses within a block  
is given by 
$W^n(\mb{y}|\mb{x}, {\theta})=\prod_{i=1}^{n}  W(y_i|x_i,\theta) $ where 
$\mb{x}=(x_1,\dots,x_n)$ and $\mb{y}=(y_1,\dots,y_n)$. 
The receiver only knows an estimate $\hat{\theta}_{R}$ of the channel state and a characterization of the estimator performance in terms of the conditional probability density function (pdf) $\psi(\theta|\hat{\theta}_{R})$ (this can be obtained using $\mc{W}_{\Theta}$, the estimator function and the a priori distribution of $\theta$). Moreover, a feedback channel provides the transmitter with noisy CSI $\hat{\theta}_{T}$ 
($\hat{\theta}_{T}$ in general is different from $\hat{\theta}_{R}$, e.g.~due to quantization).
The joint distribution of $(\hat{\theta}_{T},\hat{\theta}_{R},\theta)$ 
is given by $\psi(\hat{\theta}_{T},\hat{\theta}_R,\theta)$. The scenario underlying these assumptions is motivated by current wireless systems, where for the case of a mobile receiver $T$ may be too short to permit reliable estimation of the fading coefficients. 

The concept of outage capacity was first proposed in \cite{shamai-1994} for fading channels. It is defined as the maximum rate that can be supported with probability $1-\gamma$, where $\gamma$ is a prescribed outage probability. In contrast, ergodic capacity is the maximum information rate for which error probability decays exponentially with the code length.

In our setting, a transceiver using 
$\hat{\theta}=(\hat{\theta}_R,\hat{\theta}_T)$
instead of $\theta$ obviously might not support an information rate $R$ even if $R$ is less than the channel's capacity under perfect CSIR (even arbitrarily small rates might not be supported if $\hat{\theta}$ and $\theta$ happen to be strongly different). Consequently, outages induced by channel estimation errors will occur 
with a certain probability $\gamma$. 
The outage probability depends on the codeword error probability, averaged over a random coding ensemble and over all channel realizations given the estimated state. We first formalize the notion of estimation-induced outage capacity for general DMCs, and then we present a coding theorem providing the explicit expression for the corresponding capacity, which is a function of the outage probability $\gamma$ (Section 2). Due to the independence of 
different blocks (coherence intervals), 
it is sufficient to study the estimation-induced outage rate
$C(\gamma,\psi_{\theta|\hat{\theta}},\hat{\theta})$ for a single block (coherence interval),
for which the channel state is fixed but unknown to the transmitter and the receiver. 
Since this rate still depends on the random channel estimates $\hat{\theta}$, we
will consider the performance measure \vspace{-.1cm}
\begin{equation}
\bar{C}(\gamma,\psi_{\theta|\hat{\theta}})=E_{\hat{\theta}} \big\{ C(\gamma,\psi_{\theta|\hat{\theta}},\hat{\theta})\big\},
\label{average_capacity}
\end{equation}
which describes (average) information rate with prescribed outage probability.
The expectation in \eqref{average_capacity} is with respect to the joint distribution
$\psi(\hat{\theta})= \psi(\hat{\theta}_T| \hat{\theta}_R)\int_\Theta \psi(\hat{\theta}_R|\theta)\psi(\theta)d\theta $ and reflects an average over a large number of 	blocks (coherence intervals),
cf.~the discussion in \cite{goldsmith-1997}.

Our notion of reliably communication is relevant e.g.\ for communication systems where a quality of service (QoS) in terms of error performance must be ensured although significant channel variations occur due to user mobility. 
An example of such a scenario involving a 
fading Ricean channel with AWGN, rate-limited feedback, and maximum likelihood (ML) channel estimation, will be considered in Section 4 to illustrate the mean outage capacity $\bar{C}(\gamma,\psi_{\theta|\hat{\theta}})$.

\section{Problem Statement and Main Result}\label{sec-main-results}

In this section, we first develop a proper formalization of the notion of estimation-induced outage capacity
and state our main result. 

\subsection{Problem Definition}

%We next provide a more precise operational definition of what we mean by the outage capacity in our context, 
%i.e., the largest rate for a prescribed outage probability $\gamma$.

\def\Mcb{M}

A message $m$ from the set
$\mc{M}=\{1,\dots,\lfloor\exp(nR)\rfloor \}$ is transmitted using a 
% $\mc{M}=\{1,\dots,M_{\theta,\hat{\theta}}\}$ is transmitted using a 
length-$n$ block code defined as a pair $(\varphi,\phi)$ of mappings, % and an estimated state 
where $\varphi: \mc{M}\times  \Theta     \mapsto   \mathscr{X}^n$ is the encoder
(that utilizes $\hat{\theta}_T$), and $\phi: \mathscr{Y}^n\times  \Theta  \mapsto   \mc{M}\cup \{0\}$
is the decoder (that utilizes $\hat{\theta}_R$). The random rate, which depends on the unknown channel realization $\theta$ through its probability of error, is given by 
$n^{-1}\log \Mcb_{\theta,\hat{\theta}}$. 
%$\displaystyle{\frac{1}{n}}\log M_{\theta,\hat{\theta}} $. 
The maximum  (over all messages) error probability 
\vspace{-.1cm}
\[ e_{\max}(\varphi,\phi,\hat{\theta};\theta)=
\max_{m\in\mc{M}}\!\!\!\!\!\!\sum_{\mathbf{y}\in\mathscr{Y}^n:\phi(\mathbf{y}, \hat{\theta}_R )\neq m}\!\!\!\!\!\!\!\!\!\!\!W^n\big(\mathbf{y}|\varphi(m,\hat{\theta}_T),\theta\big).\vspace{-.1cm}
\]
% is random since the probability of error implicitly depends on the channel realization $\theta$.

For a given channel estimate $\hat{\theta}=(\hat{\theta}_R,\hat{\theta}_T)$, and $0<\epsilon,\gamma< 1$, an outage rate $R\geq 0$ is $(\epsilon,\gamma)$-achievable on an unknown channel $W(\cdot|x,\theta)\in \mc{W}_\Theta$, if for every $\delta>0$ and every sufficiently large $n$ there exists a sequence of length-$n$ block codes such that the rate satisfies\vspace{-.1cm}
$$
\Pr\Big(\big\{\theta\in \Lambda_{\epsilon}:\, n^{-1}\log \Mcb_{\theta,\hat{\theta}}\,\geq\, R-\delta \big\} \big | \hat{\theta} \Big)\geq 1-\gamma,\vspace{-.1cm}
$$
where $\Lambda_{\epsilon} = \big\{\theta\in\Theta\!: \,e_{\max}(\varphi,\phi,\hat{\theta};\theta)\leq\epsilon\big\}$ is the set of all channel states allowing for reliable decoding. 
% Via this definition, % implicitly satisfies that the 
This definition requires that
maximum error probabilities larger than $\epsilon$ occur with probability less than $\gamma$, i.e., 
$P_{\theta|\hat{\theta}}(\Lambda_{\epsilon} |\hat{\theta}) \,\geq\, 1 - \gamma.$ The practical advantage of such definition is that for any degree of accuracy channel estimation, the transmitter and receiver strive to construct codes for ensuring reliable communication with probability $1-\gamma$, no matter which unknown state $\theta$ arises during the transmission.

A rate $R\geq 0$ is $\gamma$-achievable if it is $(\epsilon,\gamma)$-achievable for every $0<\epsilon<1$. Let $C_{\epsilon}(\gamma,\psi_{\theta|\hat{\theta}},\hat{\theta})$ be the largest $(\epsilon,\gamma)$-achievable rate for an outage probability $\gamma$ and a given estimated $\hat{\theta}$. The  \emph{estimation-induced outage capacity} of this channel is then defined as the largest $\gamma$-achievable rate, i.e., 
\[
C(\gamma,\psi_{\theta|\hat{\theta}},\hat{\theta})=\lim\limits_{\epsilon\downarrow  0}C_{\epsilon}(\gamma,\psi_{\theta|\hat{\theta}},\hat{\theta}).\vspace{-4mm}
\] 

\subsection{Coding Theorem}
We next state a coding theorem quantifying the estimation-induced outage capacity $C(\gamma,\psi_{\theta|\hat{\theta}},\hat{\theta})$ 
for our scenario where an estimate $\hat{\theta}_R$ of the channel state 
% with arbitrary estimation accuracy (or estimator)
is known at the decoder and a noisy version $\hat{\theta}_T$ of $\hat{\theta}_R$ is known at the encoder.  
We impose an input constraint 
that depends on the transmitter CSI and requires that
$\Gamma(P)=\sum_{x\in \mc{X}}\Gamma(x)P(x)$
is less than 
$\mc{P} (\hat{\theta}_T)$. % \!:\!\Theta \mapsto  \mathbb{R}_{+}$.
Here, $\Gamma(\cdot)$ is any arbitrary non-negative function, and $P(\cdot)$ denotes the 
input distribution.

\begin{theorem}\label{theo-capacity}
Given $0\leq\gamma<1$ the estimation-induced outage capacity of an unknown DMC $W(\cdot|x,\theta)\in \mc{W}_\Theta$ is given by\vspace{-4mm}
\begin{equation}
\label{outage-cap}
C(\gamma,\psi_{\theta|\hat{\theta}},\hat{\theta})=\!\!\!\max\limits_{P:\, \Gamma(P)\leq \mc{P}(\hat{\theta}_T)}\mathscr{C}(\gamma,\psi_{\theta|\hat{\theta}}, \hat{\theta},P),\vspace{-5mm}
\end{equation}%\noindent
where\vspace{-1mm}
\begin{equation}
\mathscr{C}(\gamma, \psi_{\theta|\hat{\theta}},\hat{\theta},P)=\!\!\!\!\!\!\!\!\!\!\!\sup\limits_{\Lambda\subset \Theta:\,\,\Pr(\Lambda|\hat{\theta})\geq 1-\gamma}\!\!\inf\limits_{\theta\in\Lambda}I\big(P,W(\cdot|\cdot,\theta)\big).%\vspace{.5mm}
\label{qe-capacity}
\end{equation}
In addition, $C_{\epsilon}(\gamma,\psi_{\theta|\hat{\theta}},\hat{\theta})=C(\gamma,\psi_{\theta|\hat{\theta}},\hat{\theta})$ $\,\,\,\forall$ $0<\epsilon<1$.
\end{theorem}

In this theorem, we used the mutual information\vspace{-3mm}
$$
I\big(P,W(\cdot|\cdot,\theta)\big) = \sum_{x\in\mc{X}}\sum_{y\in\mc{Y}}
P(x)W(y|x,\theta)\log \frac{W(y|x,\theta)}{ Q(y|\theta) }\vspace{-3mm}
$$
with $ Q(y|\theta) = \sum_{x\in\mc{X}} P(x)W(y|x,\theta)$.
We emphasize that the supremum in \eqref{qe-capacity} is taken over all subsets $\Lambda$ of $\Theta$ 
that have (conditional) probability at least $1-\gamma$. Furthermore, codes achieving capacity \eqref{qe-capacity} can be viewed as codes for a simultaneous channel $\mc{W}_{\Lambda^*}$, which has been determined by the decoder. Hence, this outage capacity $C(\gamma,\psi_{\theta|\hat{\theta}},\hat{\theta})$ is seen to equal the maximum capacity of all compound channels that are contained in $\mc{W}_{\Theta}$ and, conditioned on $\hat{\theta}$, have sufficiently high probability. The significance of Theorem \ref{theo-capacity} is that it provides an explicit way to evaluate the outage capacity for an unknown but estimated channel for arbitrary estimation accuracies without additional assumptions. 

%A proof of Theorem \ref{theo-capacity} is needed, since the classical definition of outage capacity in terms of instantaneous mutual information cannot be used here. It requires perfect CSI which here is available neither at the transmitter nor at the receiver. 
Observe that if perfect CSIR is available then $\Lambda_\epsilon=\Theta$ and the instantaneous mutual information is attainable. Thus, every rate $R$ can be associated to the set $\Lambda_{R}=\{\theta\in\Theta:I(P,W(\cdot|\cdot,\theta))\geq R-\delta\}$ whose probability is $1-\gamma$. Therefore, in this case, the channel can be modeled as a compound channel, whose transition probability depends on a random parameter $\theta\in \Theta$. In the following section we provide a proof of Theorem \ref{theo-capacity}.

\section{Proof of the Coding Theorem}

In this section we determine the capacity by using the tools of information theory, according to the definition in Section 2. The proof of Theorem \ref{theo-capacity} is based on an extension of the maximal code lemma \cite{csiszar-book} to bound the minimum size of the images for the considered channels, according to the notion of estimation-induced outage capacity. 

Throughout this section, we will use the notion of (conditional)
information-typical (I-typical) sets defined in terms of (Kullback-Leibler) divergence, i.e., 
$\mc{T}_{P}^{n}(\delta)=\{\mathbf{x}\!\!: \mc{D}(\hat{P}_n\| P)\le \delta\}$
and
$\mc{T}_{W}^{n}(\mathbf{x},\delta)=\{\mathbf{y}\!\!: \mc{D}(\hat{W}_n\| W |\hat{P}_n)\le \delta\}$
where $\hat{P}_n$ is the empirical PM associated with $\mb{x}$
and $\hat{W}_n$ is the empirical conditional PM associated with $\mb{x}$ and 
$\mb{y}$.

\subsection{Generalized Maximal Code Lemma}\label{sub-MCL}

Let $\mathscr{I}_{\!\Lambda}$ denote the set of all common $\eta$-images $\mathscr{B}^n\subseteq \mathscr{Y}^n$ associated to a set $\mathscr{A}^n\subset \mathscr{X}^n$ via the collection of simultaneous DMCs $\mc{W}_\Lambda$,\vspace{-1mm}
$$
\mathscr{I}_{\!\Lambda} (\mathscr{A}^n\!,\eta)\!=\!\Big\{\mathscr{B}^n\!: \inf\limits_{\theta\in \Lambda} W^n(\mathscr{B}^n |\mb{x},\theta)\geq \eta\,\,\textrm{for all }\mb{x}\in \!\mathscr{A}^n \Big\}.\vspace{-1mm}
$$
In the following, we will denote by \vspace{-1mm}
\begin{equation}
\displaystyle{\textrm{g}_\Lambda (\mathscr{A}^n\!,\eta)=\!\!\!\!\min_{\mathscr{B}^n\in \mathscr{I}_{\!\Lambda} (\mathscr{A}^n\!,\eta)} \| \mathscr{B}^n\|}\vspace{-1mm}\label{min-image}
\end{equation}
the minimum of the cardinalities of all common $\eta$-images $\mathscr{B}^n$.
For a given channel estimate $\hat{\theta}=(\hat{\theta}_{_T},\hat{\theta}_{_R})$ with degraded CSIT $\theta\minuso  \hat{\theta}_{_R} \minuso \hat{\theta}_{_T}$, a code $\big(\mb{x}_1(\hat{\theta}_{_T}),\dots,\mb{x}_{M}(\hat{\theta}_{_T});\mathscr{D}^n_1(\hat{\theta}),\dots,\mathscr{D}^n_{M}(\hat{\theta})\big)$ according to the above definition consists of a set of codewords $\mb{x}_m(\hat{\theta}_{_T})$ and associated decoding sets $\mathscr{D}^n_m(\hat{\theta})$ (i.e., the decoder reads $\phi(\mathbf{y},\hat{\theta})=m$ iff $\mathbf{y}\in \mathscr{D}^n_m(\hat{\theta})$). For any set $\mathscr{A}^n$, we call a code admissible if $\mb{x}_m(\hat{\theta}_{_T})\in\mathscr{A}^n$, all decoding sets $\mathscr{D}^n_m(\hat{\theta})\subseteq  \mathscr{Y}^n$ are mutually disjoint, and the set \vspace{-1mm}
\begin{equation}
\Lambda_\epsilon=\Big\{\theta\in\Theta:\max\limits_{m\in\mc{M}}W^n\big((\mathscr{D}_m^n(\hat{\theta}))^c |\mb{x}_m(\hat{\theta}_{_T}),\theta\big) \leq \epsilon\Big\},\label{eq-MCL8}\vspace{-1mm}
\end{equation}
satisfies that $\Pr(\Lambda_\epsilon| \hat{\theta})\geq 1-\gamma$. Any input distribution satisfying the input constraint $\mc{P}(\hat{\theta}_{_T})$ is denoted by $P(\cdot|\hat{\theta}_{_T} )$.

%\vspace{1mm}

\begin{theorem}\label{theo-MCL}
Let two arbitrary numbers $0<\epsilon,\delta<1$ be given. There exists a positive integer $n_0$
such that for all $n\geq n_0$ the following two statements hold.\vspace{2mm}

1) \textit{Direct Part:} For any $\mathscr{A}^n\subset \mc{T}_{P|\hat{\theta}_{_T}}^{n}(\delta,\hat{\theta}_{_T})$ and any random set $\Lambda\subset\Theta$ with $\Pr(\Lambda| \hat{\theta})\geq 1-\gamma$, there exists an admissible sequence of length-$n$ block codes of size
\begin{equation}\label{mcl-lower}
M_{\theta,\hat{\theta}} \geq \exp\big[-n\big(H(\mc{W}_\Lambda | P) -\delta  \big) \big]
     \textrm{g}_\Lambda  (\mathscr{A}^n,\epsilon -\delta),
\end{equation}
for all $\theta\in\Lambda$, where $\Lambda_\epsilon=\Lambda$.

2) \textit{Converse Part:} 
For $\mathscr{A}^n = \mc{T}_{P|\hat{\theta}_T}^{n}(\delta,\hat{\theta}_{_T})$, the size of any admissible sequence of length-$n$ block codes is bounded 
\begin{equation}\label{mcl-upper}
M_{\theta,\hat{\theta}}   \leq \exp\big[-n\big(H(\mc{W}_{\Lambda_\epsilon} | P) +\delta  \big) \big]    \textrm{g}_{\Lambda_\epsilon} (\mathscr{A}^n,\epsilon +\delta),
\end{equation}
for all $\theta\in\Lambda_\epsilon$.
\end{theorem}

The proof of this theorem easily follows from basic properties of I-typical sequences and the concept of robust I-typical sets in Appendix \ref{appendices}. Whereas, Theorem \ref{theo-capacity} is obtained through the following corollary. 

\begin{corollary}
For a given channel estimate $\hat{\theta}$ and an outage probability $\gamma$, and $0<\epsilon,\delta<1$ and any PM $P(\cdot|\hat{\theta}_{_T})\in \mc{P}(\mathscr{X})$. Let $\mathscr{C}(\gamma,\psi_{\theta|\hat{\theta}}, \hat{\theta},P)$ be defined by expression \eqref{qe-capacity}. Then the following statements holds: 

(i) There exists an optimal sequence of block codes of length $n$ and size $M_{\theta,\hat{\theta}}$, whose maximum error probabilities larger than $\epsilon$ occur with probability less than $\gamma$, such that\vspace{-1mm}
\begin{equation}
\Pr\Big(n^{-1}\log M_{\theta,\hat{\theta}} \geq R-2\delta \big |\hat{\theta} \Big)\geq 1-\gamma \label{eq-lower}
\vspace{-1mm}
\end{equation}
for all rate $R \leq \mathscr{C}(\gamma,\psi_{\theta|\hat{\theta}}, \hat{\theta},P)$, provided that $n\geq n_0$. 

(ii) For any block codes of length $n$, size $M_{\theta,\hat{\theta}}$ and codewords in $\mc{T}_{P|\hat{\theta}_{_T}}^{n}(\delta,\hat{\theta})$, whose maximum error probabilities larger than $\epsilon$ occur with probability less than $\gamma$. The largest code size satisfies\vspace{-1mm}
\begin{equation}
 \Pr\Big(n^{-1}\log M_{\theta,\hat{\theta}} > R+2\delta \big |\hat{\theta} \Big)<\gamma\label{eq-upper}\vspace{-1mm}
\end{equation}
for all rate $R \geq \mathscr{C}(\gamma, \psi_{\theta|\hat{\theta}},\hat{\theta},P)$, whenever $n\geq n_0$.
\end{corollary}

\emph{Proof:} From the direct part of Theorem \ref{theo-MCL} and Lemma \ref{lemma-higher-prob}, we have that there exists admissible codes such that \vspace{-1mm}
\begin{equation}
n^{-1} \log M_{\theta,\hat{\theta}} \geq n^{-1}  \log\textrm{g}_\Lambda \big(\mathscr{A}^n,\epsilon -\delta\big)-H(\mc{W}_\Lambda | P)-\delta,  \label{eq-MCLC-1}
\end{equation}
for all $\theta\in \Lambda$ and sets $\Lambda\subset \Theta$ (having probability at least $1-\gamma$). Let
$\hat{ \mathscr{D}}^n$ be the common $(\epsilon-\delta)$-image of minimal size
$\|\hat{\mathscr{D}}^n\|=\textrm{g}_\Lambda \big(\mathscr{A}^n,\epsilon -\delta\big)$.
Then it is easy to show that $\inf\limits_{\theta\in \Lambda} W_\theta P^n(\hat{\mathscr{D}}^n) \geq (\epsilon-\delta)^2$. By applying Corollary 1.2.14 in \cite{csiszar-book} to this relation and 
substituting it in (\ref{eq-MCLC-1}), we obtain for all $n\geq n_0^\prime(| \mathscr{X}|,|\mathscr{Y} |,\epsilon,\delta)$, \vspace{-1mm}
\begin{eqnarray}
n^{-1} \log M_{\theta,\hat{\theta}} &\geq&   \inf\limits_{\theta\in \Lambda}I(P, W(\cdot|\cdot,\theta )) -2\delta, \label{eq-MCLC-3} \vspace{-1mm}
\end{eqnarray}
for all $\theta\in\Lambda$, where the last inequality follows from the concavity 
of the entropy function with respect to
$W_\theta$. Finally, taking the supremum in \eqref{eq-MCLC-3} with respect to all sets $\Lambda\subset\Theta$ having probability at least $1-\gamma$ yields the lower bound (\ref{eq-lower}) \vspace{-1mm}
\begin{equation}
n^{-1} \log M_{\theta,\hat{\theta}}  \geq \mathscr{C}(\gamma,\psi_{\theta|\hat{\theta}}, \hat{\theta},P)  -2\delta \geq R-2\delta,\vspace{-1mm}
\end{equation}
for all rate $R \leq \mathscr{C}(\gamma, \psi_{\theta|\hat{\theta}},\hat{\theta},P)$ and $\theta\in\Lambda^*$, which is attained by some code with $\Lambda_\epsilon=\Lambda^*$. Next we prove the upper bound (\ref{eq-upper}). 
From the converse part of Theorem \ref{theo-MCL}, we have \vspace{-1mm}
\begin{equation}
n^{-1} \log M_{\theta,\hat{\theta}}  \leq n^{-1} \log\textrm{g}_{\Lambda_\epsilon} \big(\mathscr{A}^n,\epsilon +\delta\big) -H(\mc{W}_{\Lambda_\epsilon} | P) +\delta, \label{eq-MCLC-10}
\end{equation}
for all $\theta\in\Lambda_\epsilon$. Since $\mathscr{A}^n= \mc{T}_{P|\hat{\theta}_{T}}^{n}(\delta,\hat{\theta})$ implies that any common $(\epsilon+\delta)$-image of $\mathscr{A}^n$ will be included in $\bigcap\limits_{\theta \in \Lambda_\epsilon}\mc{T}_{W_\theta P}^{n}(\delta_n^\prime)$, Lemma 1.2.12 in \cite{csiszar-book} ensures that there 
exists $n\geq n_0^{\prime\prime}(| \mathscr{X}|,|\mathscr{Y} |,\epsilon,\delta)$ such that, \vspace{-1mm}
\begin{equation}
  n^{-1} \log\textrm{g}_{\Lambda_\epsilon}\big(\mathscr{A}^n,\epsilon +\delta\big) \leq \inf\limits_{\theta\in \Lambda_\epsilon} H(W_\theta P)+\delta.   \label{eq-MCLC-4} \vspace{-2mm}
\end{equation}  
Then by applying equation (\ref{eq-MCLC-4}) to equation (\ref{eq-MCLC-10}), and then by taking its supremum with respect to all sets $ \Lambda \subset \Theta$ having probability at least $1-\gamma$, we obtain\vspace{-1mm}
\begin{equation}
n^{-1} \log M_{\theta,\hat{\theta}} \leq \mathscr{C}(\gamma, \psi_{\theta|\hat{\theta}},\hat{\theta},P) +2\delta \leq  R +2\delta,\vspace{-1mm}
 \label{eq-MCLC-5}
\end{equation}
for all $R\geq  \mathscr{C}(\gamma,\psi_{\theta|\hat{\theta}}, \hat{\theta},P)$ and $\theta\in \Lambda_\epsilon$ with $\Pr(\theta\notin\Lambda_\epsilon | \hat{\theta})< \gamma$, and this concludes the proof.

\section{Numerical Results and Discussion}\label{sec-aplications}

In this section, we illustrate our results via a realistic 
single user mobile wireless communication system involving a 
Ricean block flat fading channel, where the channel state is described by
a single fading coefficient. The channel states in each block are i.i.d.~and
unknown at the transmitter and the receiver. The transmission extends over many blocks (coherence intervals)
such that the average outage capacity \eqref{average_capacity} is indeed the appropriate
performance criterion.
The practical significance of this capacity stems from QoS requirements present in many communication services.

Within each block, the actual codeword (data) is preceded by a length-$N$ training sequence 
$\mathbf{x}_T=[x_0,\dots,x_{N-1}]$ of power $P_T$ which is known by the receiver. This enables maximum likelihood (ML) channel estimation of the fading coefficient $\theta$ at the receiver yielding the estimate $\hat{\theta}_{R}$.
In many wireless systems, CSI at the transmitter $\hat{\theta}_{T}$ has to be provided by the receiver via
a feedback/CSIT. 
This allows the transmitter to perform power control $\mc{P}(\hat{\theta}_{R})$, i.e., allocate more transmit power when the estimated channel is good, and less or no power when the channel is bad. Below, we consider the following three feedback schemes: 
(i) no feedback, i.e., absence of CSIT; (ii) an instantaneous and unlimited feedback/CSIT  ($\hat{\theta}_T=\hat{\theta}_R$); (iii) an instantaneous and rate-limited feedback/CSIT; here the CSI is quantized using a quantization codebook which is known at the transmitter and the receiver (we construct this codebook using the well-known Lloyd-Max algorithm). 

\subsection{Channel Model} The channel model within a block is given by (all quantities are complex-valued) 
$Y[i]=H[i]\,X[i]+Z[i]$, where 
$X[i]$ and
$Y[i]$ are the discrete-time transmit and receive signal, respectively, $H$ is the fading coefficient, and $Z[i]\sim \mc{CN}(0,\sigma^2_Z)$ is 
i.i.d.~zero-mean, circularly complex Gaussian noise. The transmit signal is subject to the average power constraint $E\big\{|X[i]|^2\big\}\leq \mc{P}(\hat{\theta}_T)$ with $E_{\hat{\theta}_T}\big\{\mc{P}(\hat{\theta}_T)\big\}\leq P$. The optimum power allocation $\mc{P}(\hat{\theta}_{R})$ is obtained using Lagrange multipliers and the Kuhn-Tucker theorem. The channel state $\theta=H[i]$ is assumed to be circularly complex Gaussian $\theta \sim \psi(\theta)=\mc{CN}\big(\mu_h, 2\sigma^2_h\big)$. The Rice factor is defined as $K_h=\frac{|\mu_h|^2}{2\sigma^2_h}$. 
The ML estimate
$\hat{\theta}_{R}=\hat{H}[i]$ is obtained by correlating the received signal with the 
known training sequence $\mathbf{x}_T$.
Its performance can be characterized via the %conditional 
pdf $\psi(\theta | \hat{\theta}_R)= \mc{CN}\big(\rho \hat{\theta}_R +(1-\rho) \mu_h ,  \rho \sigma_W^2 \big)$, where $\rho=\frac{2 \sigma_h^2  }{\sigma_W^2+ 2\sigma_h^2 }$ and $\sigma_W^2=\sigma_Z^2/(N P_T)$. 

For a given estimate $\hat{\theta}_{0}$, to evaluate \eqref{qe-capacity} requires
solving an optimization problem where we have to determine the optimum set $\Lambda^*$, and the associated channel state $\theta^*\in \Lambda^*$ minimizing mutual information. The estimation-induced outage capacity can then be shown to be given by   %\vspace{-1mm}
$
C(\gamma,\psi_{\theta|\hat{\theta}},\hat{\theta}_0)=\log_2\!\left(1+\frac{\big(r^*(\gamma,\hat{\theta}_{0})\big)^2 \mc{P}(\hat{\theta}_{_T,0})}{\sigma_Z^2}\right),
$
where $r^*$ is the $\gamma$-percentile\footnote{It can be computed by using the cumulative distribution of a non-central chi-square of two degrees of freedom.} of $\psi\big(r|\hat{\theta}=\hat{\theta}_0\big)$ with $r=|\theta|$ (for further details see \cite{spawc06}).

\subsection{Results and Discussion}

\begin{figure}[ht]
\centering
\includegraphics[width=2.7in]{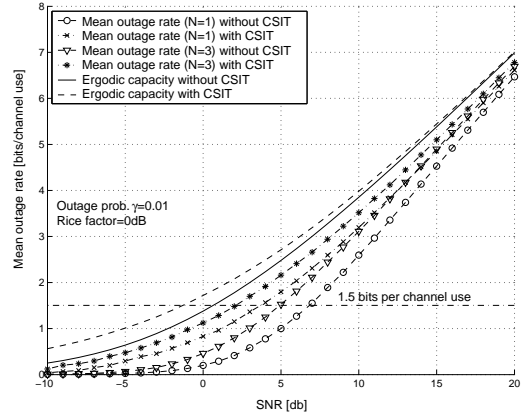}
\vspace{-.2cm}\caption{Average estimation-induced outage capacity for different amounts of training vs.~SNR.}\vspace{-.6cm}
\label{fig}
\end{figure}

% In this extended abstract, numerical results based on Monte Carlo simulations are presented for cases (i) and (iii) above. 
% Optimal power allocation policities $\mc{P}(\hat{\theta}_T)$ are obtained by using Lagrange multipliers and the Kuhn-Tucker theorem. 
Fig.\ \ref{fig} shows the average estimation-induced outage capacity (cf.~\eqref{average_capacity}) in bits per channel use for outage probability $\gamma=0.01$ versus the signal-to-noise ratio $\text{SNR}=|\mu_h|^2P/ \sigma_Z^2$
for different amounts of training and for unlimited and absent feedback/CSIT
(all numerical results were obtained using Monte Carlo simulations).
For comparison, we show ergodic capacity under perfect CSI. The channel's Rice factor was $K_h=0\,$dB. It is seen that the average rate increases with the amount of CSIR and CSIT.
To achieve 1.5 bit per channel use without feedback/CSIT, it is seen that 
a scheme with estimated CSIR and $N=3$ ($\nabla$ markers)
requires $5\,$dB, i.e., 4.3\,dB more than with perfect CSIR (solid line). Whereas if the training
 length is further reduced to $N=1$ ($\circ$ markers), this gap increases to 6.4\,dB. In the case of unlimited feedback (CSIT=CSIT), the SNR requirements for
1.5 bit per channel use are
$-1.3\,$dB (perfect CSIR, dashed line),
$2.1\,$dB (estimated CSIR with $N\!=\!3$, $\ast$ markers),
and
$3.7\,$dB (estimated CSIR with $N\!=\!1$, $\times$ markers), respectively.
Thus, with unlimited feedback the gap between estimated and perfect CSI is slightly smaller than
without feedback (3.4\,dB and 5\,dB with $N\!=\!3$ and $N\!=\!1$, respectively). Observe that for values of SNR larger than $10\,$dB similar performance are achieved without feedback/CSIT and $N=3$ comparing to a system with unlimited feedback and $N=1$. Therefore, using this information a system designer may decide to use training sequences of length $N=3$ instead of implementing a feedback channel.

\begin{figure}[ht]
\centering
\includegraphics[width=2.7in]{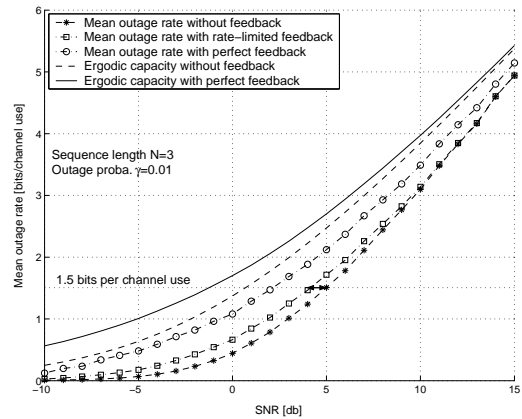}
\vspace{-.2cm}\caption{Average estimation-induced outage capacity with two bits of rate-limited feedback ($R_{FB}=2$) vs.~SNR.\vspace{-.2cm}}
\label{rate-limited}
\end{figure}

Fig.\ \ref{rate-limited} shows the average estimation-induced outage capacity for an outage probability $\gamma=0.01$ and rate-limited feedback/CSIT versus the SNR. We suppose two bits of feedback ($R_{RF}=2$), and training sequences of length $N=3$. Observe that at $1.5\,$bits the gap between the average outage capacity without feedback and rate-limited feedback is $1\,$dB for two bits of feedback/CSIT. Whereas the gap respect to the average outage capacity with unlimited feedback is only $2\,$dB.   \vspace{-1mm}

\section{Conclusions}

In this paper we have studied the problem of reliable communications over unknown DMCs when the receiver and transmitter only know an estimate of the channel state. We proposed to characterize the information theoretic limits of such scenarios in terms of the novel notion of estimation-induced outage capacity. We provided an explicit expression for the maximum achievable outage rate in the context of an associated coding theorem and its strong converse. We used a Ricean fading channel and maximum likehood channel estimation to illustrate our approach by computing its mean outage capacity. Our results are useful to assess the amount of training data and feedback required to achieve a target rate satisfying  a quality of service constraint. It will be attractive to study coding schemes achieving this capacity because this allows to design communication systems with QoS constraints and imperfect channel estimation.\vspace{-1mm}

\section*{Acknowledgment}
The authors are grateful to Prof.\ Te Sun Han for many helpful discussions and suggestions on the technical aspects of the paper's proofs.\vspace{-1mm}

\appendix
\section{Auxiliary results}\label{appendices}


This appendix introduces a few concepts and furthermore provides some auxiliary technical results required for the proof of Theorem \ref{theo-capacity}.

\emph{Robust Decoders:} Let $\mathscr{A}^n\subset \mathscr{X}^n$ denote a set of 
transmit sequences and let $W_\theta(\cdot|x)=W(\cdot|x,\theta)$.
A set $\mathscr{B}^n\subset \mathscr{Y}^n$ (depending on $\Lambda\subset\Theta$) 
is called a {\em robust $\epsilon$-decoding set} for a sequence $\mb{x}\in \mathscr{A}^n$ 
and an unknown DMC $W(\cdot|x,\theta)\in \mc{W}_\Theta$, 
if $\Pr \big( W^n(\mathscr{B}^n |\mb{x},\theta)>1-\epsilon\big|  \hat{\theta}\big)\geq 1-\gamma$.
 
A set $\mathscr{B}^n\subset \mathscr{Y}^n$ of receive sequences is called a 
{\em common $\eta$-image} ($0<\eta\leq 1$) of a transmit set $\mathscr{A}^n\subset \mathscr{X}^n$ for 
the collection of DMCs
$\mc{W}_\Lambda$, iff $ \inf\limits_{\theta\in \Lambda} W^n(\mathscr{B}^n |\mb{x},\theta)\geq \eta $ 
for all $\mb{x}\in \mathscr{A}^n$.
Finally, $\Lambda\subset\Theta$ is called a {\em confidence set} for $\theta$ given $\hat\theta$,
if $\Pr(\theta \notin \Lambda|  \hat{\theta})<\gamma$ and $\gamma$ is the outage probability.

\emph{Robust I-Typical Sets:} A robust I-typical set is defined as 
$
\mathscr{B}^n_\Lambda(\mb{x},\delta_n)
=
{\bigcup\limits_{\theta\in \Lambda}}\mc{T}_{W_\theta}^{n}(\mathbf{x},\delta_n),
$
with arbitrary $\Lambda \subset \Theta$ and $\delta$-sequence $\{\delta_n\}$ (cf.\ \cite{csiszar-book}).

\vspace{-2mm}
\begin{lemma}\label{lemma-higher-prob}
For any $0<\gamma,\epsilon< 1$, 
a necessary and sufficient condition for a robust I-typical set $\mathscr{B}^n_\Lambda({\bf x},\delta_n)$ to be 
a robust $\epsilon$-decoding set with probability $1-\gamma$ is that $\Lambda$ be a confidence set. \vspace{-2mm}
\end{lemma}

%\emph{Proof:} It is straightforward to show that $\mathscr{B}^n_\Lambda({\bf x},\theta)$ is a common $\eta$-image for the collection of DMCs $\mc{W}_\Lambda$ with $\eta=1-\epsilon$. Hence, the necessity is a direct consequence of Proposition \ref{prop-conf-set}. The sufficiency follows from basic properties of I-typical sequences.

\begin{theorem}\label{theo-cardinality}
For any collection of DMCs $\mc{W}_\Lambda$ and associated robust I-typical set 
$\mathscr{B}_{\Lambda}^n(\mathbf{x},\delta_n)$ with
$\mathbf{x} \in \mc{T}_{P}^{n}(\mb{x}, \delta_n)$, there exists an index $n_0$ such that for all $n\geq n_0$ the size 
$\|\mathscr{B}^n_\Lambda(\mathbf{x}, \delta_n)\|$
of the  robust I-typical set is bounded as \vspace{-2mm}
\[ 
 \Big| \frac{1}{n}\log \|\mathscr{B}^n_\Lambda(\mathbf{x}, \delta_n)\|  - H(\mc{W}_\Lambda | P) \Big| \leq  \eta_n.  \vspace{-2mm}
\] 
Here, $H(\mc{W}_\Lambda | P)=\sup\limits_{V \in\mc{W}_\Lambda }H(V | P)$ and 
$ \eta_n  \rightarrow 0$ as $\delta_n  \rightarrow 0$ and $n \rightarrow \infty$.\vspace{-2mm}
\end{theorem}

\emph{Proof:} We first show that the size of $ \mathscr{B}_\Lambda^n(\mathbf{x},\delta_n) $ is asymptotically
equal to the size of 
$\mathscr{B}_\Sigma^n(\mathbf{x})= \bigcup\limits_{V\in\Sigma }\mc{T}_{V}^{n}(\mathbf{x})$
where $\Sigma=\mc{W}_\Lambda\cap\mc{P}_n(\mathscr{Y})$
is the intersection of $\mc{W}_\Lambda$ with the set $\mc{P}_n(\mathscr{Y})$ of empirical 
distributions induced by receive sequences of length $n$.
In particular, there exists an index $n_0$ such that for all $n\geq n_0$ and $\mb{x}\in \mathscr{X}^n$\vspace{-1mm}
\begin{eqnarray} 
\|\mathscr{B}_\Sigma^n(\mathbf{x})\| \leq  \| \mathscr{B}_\Lambda^n(\mathbf{x},\delta_n)\|  \leq (1+n)^{|\mathscr{X}\|\mathscr{Y}|} \|\mathscr{B}_\Sigma^n(\mathbf{x})\|.\vspace{-2mm}
\label{eq-bounds-B}
\end{eqnarray}
The lower bound in \eqref{eq-bounds-B} is trivial. We will next establish that 
there exists $\epsilon_n>0$ such that for all $n\geq n_0$\vspace{-2mm}
\begin{equation} 
\bigcup\limits_{W \in \mc{W}_\Lambda} \mc{T}_{W}^{n}(\mathbf{x},\delta_n) 
\subseteq 
\bigcup\limits_{V \in \Sigma} \mc{T}_{V}^{n}(\mathbf{x},\epsilon_n),\vspace{-2mm}
\label{eq-inclusion}
\end{equation} 
from which the upper bound in \eqref{eq-bounds-B} follows via
basic properties of types (cf.\ \cite{csiszar-book}).

Assume that $\mc{W}_\Lambda$ is a relatively $\tau_0$-open subset of $\mc{W}_\Lambda \cup \mc{P}_n(\mathscr{Y})$, i.e., every $W\in \mc{W}_\Lambda$ has a $\tau_0$-neighborhood defined in the $\tau_0$-topology \cite{csiszar-sanov-1984}. Then there exists $n_0$ such that for any $n\geq n_0$ and $\varepsilon >0$, the $\varepsilon $-open ball 
$U_0(W,\varepsilon)$ satisfies
$U_0(W,\varepsilon) \cap \mc{P}_n(\mathscr{Y}) \subset \mc{W}_\Lambda$. Choose $0<\varepsilon^\prime  <\varepsilon$ and pick an empirical conditional PM $V\in\mc{P}_n(\mathscr{Y})$ such that for all 
$(a,b)\in \mathscr{X}\times\mathscr{Y}$,
$| V(b|a)- W(b|a)|<\varepsilon^\prime_n$ and $V(b|a)=0$ if $W(b|a)=0$.  
The continuity properties of information divergences imply that for
any sequence $\mb{y}\in  \mc{T}_{W}^{n}(\mathbf{x},\delta_n)$ (i.e., $\mc{D}(\hat{W}_n\| W | \hat{P}_n)  \leq \delta_n$), $ \big |\hat{W}_n(b|a)\hat{P}_n(a)  - W(b|a)\hat{P}_n(a)  \big | \leq \sqrt{\delta_n/2}$ and hence $\big |\hat{W}_n(b|a)\hat{P}_n(a) - V(b|a)\hat{P}_n(a)  \big | \leq \varepsilon^\prime  +\sqrt{\delta_n/2}$. 
Finally, from this equation it is easy to show that there exists an $\epsilon_n>0$ such that 
$\mc{D}(\hat{W}_n\|V | \hat{P}_n)\leq\epsilon_n$, i.e., $\mb{y}\in  \mc{T}_{V}^{n}(\mathbf{x},\epsilon_n)$. Consequently, for any $W\in \mc{W}_\Lambda$ and large enough $n$, it is possible to find 
$V \in \Sigma$ and $\epsilon_n >0$ such that $ \mc{T}_{W}^{n}(\mathbf{x},\delta_n)\subseteq  \mc{T}_{V }^{n}(\mathbf{x},\epsilon_n)$, thus establishing \eqref{eq-inclusion}. Using similar arguments as above and the uniform continuity of the entropy function,
it can be shown that there exists $n_0^\prime$ such that for all $n\geq n_0^\prime$ and $\mb{x}\in \mathscr{X}^n$\vspace{-2mm}
\begin{equation} 
 \Big|  \frac{1}{n}\log\|\mathscr{B}_\Sigma^n(\mathbf{x})\| - \sup\limits_{V\in \mc{W}_\Lambda}H(V|P)   \Big| \leq \xi_n,\vspace{-2mm}
\label{eq-faltante}
\end{equation} 
with $\xi_n= |\mathscr{X}||\mathscr{Y}|n^{-1}\log (n\!+\!1) +\xi_n^\prime$ and $\xi_n^\prime \rightarrow 0$ as 
$n \rightarrow \infty$. The theorem follows by combining the inequalities \eqref{eq-bounds-B} and \eqref{eq-faltante} and setting $\eta_n=\xi_n+ |\mathscr{X}||\mathscr{Y}|n^{-1}\log (n\!+\!1)$. 

\emph{ Proof of Theorem \ref{theo-MCL}}: 
To prove the direct part, consider an admissible code that is maximal, i.e., it cannot be extended by arbitrary $(\mb{x}_{M+1};\mathscr{D}_{M+1}^n)$ such that the extended code remains admissible. Define the set $ \mathscr{D}^n= \bigcup_{i=1}^{M} \mathscr{D}^n_i$ with $\mathscr{D}^n_i\subseteq \mathscr{B}^n_\Lambda(\mb{x}_i,\delta)$, and choose $\delta<\epsilon$ such that $1-\epsilon>\epsilon-\delta$. Then \vspace{-4mm}
\begin{equation}
\inf\limits_{\theta\in \Lambda} W^n(\mathscr{D}^n|\mb{x}_i,\theta)> \epsilon -\delta,\vspace{-2mm}
\label{eq-MCL1}
\end{equation}
for all $\mb{x}_i\in \mathscr{A}^n$. As the code is maximal, for all $\mb{x}\in \mathscr{A}^n\setminus \big\{\mb{x}_1,\dots,\mb{x}_{M}\big\}$, we have 
$
\inf\limits_{\theta\in \Lambda} W^n(\mathscr{B}^n_\Lambda \setminus  \mathscr{D}^n |\mb{x},\theta)\leq 1-\epsilon.
$
This equation implies that for all $\theta\in \Lambda$ and large enough $n$\vspace{-1mm}
\begin{equation}
W^n(\mathscr{D}^n|\mb{x},\theta)\geq \epsilon-\delta, \vspace{-1mm}
\label{eq-MCL3}
\end{equation}
for all $\mb{x}\in \mathscr{A}^n\setminus \big\{\mb{x}_1,\dots,\mb{x}_{M}\big\}$. The inequalities \eqref{eq-MCL1} and (\ref{eq-MCL3}) together imply that
$\mathscr{D}^n$ is a common $(\epsilon-\delta)$-image of the set $ \mathscr{A}^n$ via the
collection of channels $\mc{W}_\Lambda$. By the definition of $\textrm{g}_\Lambda (\mathscr{A}^n,\epsilon-\delta)$
it follows that %\vspace{-1mm}
\begin{equation}
\|\mathscr{D}^n\|\geq \textrm{g}_\Lambda (\mathscr{A}^n,\epsilon-\delta).%\vspace{-1mm}
\label{eq-MCL4}
\end{equation}
On the other hand, $\mathscr{D}_i^n\subseteq \mathscr{B}^n_\Lambda(\mb{x}_i,\delta)$ implies that%\vspace{-1mm}
\begin{equation}
\|  \mathscr{D}^n\| \leq  M_{\theta,\hat{\theta}}  \exp\big[n\big(H(\mc{W}_\Lambda | P) +\delta  \big) \big],\label{eq-MCL5}%\vspace{-1mm}
\end{equation}
for $n$ large enough and all $\theta\in\Lambda$, where the last inequality follows by applying the cardinality upper bound of Theorem \ref{theo-cardinality}. 
The lower bound \eqref{mcl-lower} is then immediately obtained by combining
(\ref{eq-MCL4}) and  (\ref{eq-MCL5}). 
To prove the converse part,
let $\hat{ \mathscr{D}}^n$ be a common $(\epsilon+\delta)$-image via the collection of channels $\mc{W}_{\Lambda_\epsilon}$, i.e.,\vspace{-1mm}
\begin{equation}
\inf\limits_{\theta\in \Lambda_\epsilon} W^n(\hat{ \mathscr{D}}^n |\mb{x}_m,\theta)\geq \epsilon+\delta,\,\quad \textrm{for } m\in\mc{M},\vspace{-1mm}
\label{eq-MCL6}
\end{equation}
that achieves the minimum in \eqref{min-image}, i.e.,  
$\|\hat{ \mathscr{D}}^n\|= \textrm{g}_{\Lambda_\epsilon} (\mathscr{A}^n,\epsilon+\delta)$. For any admissible code, (\ref{eq-MCL8}) and (\ref{eq-MCL6}) imply 
$
\inf\limits_{\theta\in\Lambda_\epsilon} W^n(\mathscr{D}_m^n \cap \hat{ \mathscr{D}^n}|\mb{x}_m,\theta)\ge \delta\,\quad \textrm{for } m\in\mc{M}.
$
Using Corollary 1.2.14 in \cite{csiszar-book}, we hence obtain \vspace{-1mm}
\begin{eqnarray}
\big\|\mathscr{D}_m^n \cap \hat{ \mathscr{D}}\big\| \geq \exp\big[ n\big( H(\mc{W}_{\Lambda_\epsilon} | P) -\delta \big) \big], \label{eq-MCL9}\vspace{-1mm}
\end{eqnarray}
for $n$ large enough. On the other hand, the decoding sets $\mathscr{D}_m^n$ are disjoint and thus\vspace{-1mm}
\begin{equation*}
 \textrm{g}_{\Lambda_\epsilon} (\mathscr{A}^n,\epsilon+\delta)
 = \|\hat{ \mathscr{D}}^n\| \geq  M_{\theta,\hat{\theta}} \exp\big[ n\big(H(\mc{W}_{\Lambda_\epsilon} | P) -\delta\big)\big],\vspace{-1mm}
\end{equation*}
where the last inequality follows from (\ref{eq-MCL9}). This inequality is
equivalent to \eqref{mcl-upper} and concludes the proof of the theorem.

\vspace{-.2cm}

\footnotesize
\bibliographystyle{IEEEtran.bst}
\bibliography{../../bibliography/biblio}

\end{document}